\documentclass[a4paper,11pt]{article}

\usepackage[utf8]{inputenc}
\usepackage[T1]{fontenc}
\usepackage[english]{babel}
\usepackage[hyphens]{url}
\usepackage[pdftex,urlcolor=black,colorlinks=true,linkcolor=black,citecolor=black]{hyperref}
\usepackage{csquotes}
\usepackage{fourier}
\usepackage[top=3cm,left=3.5cm,right=3.5cm,bottom=4cm]{geometry}
\usepackage{array}
\usepackage{multirow}
\usepackage{float}
\usepackage{hyperref}
\usepackage{graphicx}
\usepackage{natbib}
\usepackage{array}

\usepackage{color}

\newcounter{example}

\usepackage{fancyvrb}

\usepackage{amsmath}

\usepackage{bm}

\title{How hot is .brussels? \\
Analysis of the uptake of the .brussels\\
top-level domain name extension}

\author{Margot Waty,\thanks{Corresponding author} Seth van Hooland, Simon Hengchen,\\
Mathias Coeckelbergs and Max De Wilde \\ \\
Université libre de Bruxelles (ULB)\\
ReSIC Research Center\\ 
Information and Communication Science Department\\ 
Avenue F.D.\,Roosevelt\,50 -- CP\,123 --
B-1050 Brussels, Belgium\\
\texttt{\{margwaty,svhoolan,shengche,mcoeckel,madewild\}@ulb.ac.be}
}

\date{}

\begin{document}
\maketitle

\begin{abstract}
The opening up of the top-level domain name market in 2012 has offered new perspectives
for companies, administrations and individuals to include a geographic component within the domain name of their website.
Little to no research has been carried out since then to analyse the uptake of the new top-level domain names (TLDN). 
Based on the specific case of the TLDN .brussels, this article proposes an empirical study of how
the opening up of the top-level domain name market actually impacts registration practices. 
By making use of freely available software tools such as
OpenRefine and Natural Language Processing (NLP) methods, the entire corpus of the .brussels domain names (6300) was analysed from
a quantitative perspective. Based on a statistically representative sample (592 domain names), a qualitative interpretation allowed a more
fine-grained analysis of how the new TLDN is being used in practice. By doing so, the article gives a detailed insight into the impact of the recent
changes of the rules concerning domain name registration. Researchers, policy makers, investors and anyone who cares about
the Brussels identity in the digital realm can gain through this analysis a better understanding of the state of play of the .brussels TLDN.
\end{abstract}

\section{Introduction}
Real estate professionals often say there are three important parameters to take into account when purchasing a house or apartment: the location, the location and the location. We could not agree more in the context of the Web: domain names are the cornerstone of our digital economy . The analogy with the housing market does not stop there. Just as you permanently need to invest in a house (from minor maintenance such as polishing wooden floors to major renovation of the roof, for example), the investment in a Uniform Resource Locator (URL) does not stop at acquiring a domain name. Commercial parties are very much aware of the importance of URLs, illustrated by practices such as the domain name after market where URLs are bought and sold by commercial parties. The gradual evolution to a more structured and Semantic Web, combined with the Internet of Things,  will only emphasise the importance of stable and meaningful domain names \cite{vanhooland}.

This article proposes an in-depth study of URLs with the TLDN .brussels.
Why is the particular case of .brussels interesting to analyse? Compared to many of the other newly created TLDN,
such as \texttt{.music} or \texttt{.photography}, the range of usage is potentially a lot more diverse. Even when
compared with other geographic TLDN which reflect city names, such as \texttt{.gent} for example, the eight letter
term \enquote{Brussels} is being used in a large variety of contexts. The disambiguation page on Wikipedia
gives us an overview of the different meanings:\footnote{\url{https://en.wikipedia.org/wiki/Brussels_(disambiguation)}}

\begin{itemize}
\item the city of Brussels
\item the region of Brussels
\item the capital of Belgium, the Flemish and the French Community 
\item the metonym for the different institutions of the European Union (EU) and the European Commission (EC) more in particular
\item five different cities spread out over Canada and the United States 
\end{itemize}

This ambiguity is put into light in the context of Linked Open Data (LOD) \cite{these_max}.
The resource for Brussels is well represented in LOD datasets.
Its DBpedia page\footnote{\url{http://dbpedia.org/page/Brussels}} lists 128 equivalent resources in various knowledge bases, from Freebase to LinkedGeoData and to Global Administrative Areas (GADM).
Among those, we can find two URLs from GeoNames: \url{http://sws.geonames.org/2800866/} and \url{http://sws.geonames.org/2800867/}.
The former corresponds to Brussels as the capital of a political entity and the latter to Brussels Capital as a first-order administrative division.
The first has a population of 1~019~022, the second of 1~830~000.
So how can these two URLs \emph{actually refer to the same thing}?

The April 2016 terror attacks and Donald Trump's \enquote{[Brussels] is like living in a hellhole}\footnote{As reported by \href{http://uk.businessinsider.com/donald-trump-brussels-muslim-ban-hellhole-2016-1}{Business Insider}: \enquote{You go to Brussels -- I was in Brussels a long time ago, 20 years ago, so beautiful, everything is so beautiful -- it's like living in a hellhole right now}. The day after these comments, a citizen reacted by launching \url{http://hellhole.brussels} to counter attack.} quote also demonstrate how rapidly the semantics associated to a term can result in very different concepts. 
\citet[p. 60]{Frege84} draws a distinction between the idea, the sense, and the reference of an object:

\begin{quote}
The reference of a proper name is the object itself which we designate by its means; the idea, which we have in that case, is wholly subjective; in between lies the sense, which is indeed no longer subjective like the idea, but is yet not the object itself.
\end{quote}

Even when the reference of Brussels is agreed upon (e.g. the capital city and not the European institutions), the idea Trump has of Brussels is wildly different of that of the average city dweller.
The sense put into the concept of Brussels can therefore vary quite radically, allowing for a whole scale of interpretations.
We hope that these examples demonstrate the richness but also the complexity of the particular .brussels TLDN across its potential uses.

After a short introduction to the intriguing world of domain name registration, the article will introduce the case-study and detail which methods have been used to perform both the quantitative and qualitative analyses.
The article ends with a discussion of the results.  

\section{A rough guide to the domain name market place}

\subsection{Role of the Domain Name System (DNS)}
To better understand the role of the DNS on the Internet and the impact 
that domain names have on the image that we build online, a slight detour to its foundations can be useful. 
The core principle of the Internet is to offer a consistent and standardised platform so that different entities can quickly and easily 
communicate with each other, and share some data. To find something online, you need to be able to identify and locate the wished information. 
To do so, you will need common communication procedures like the use of a common language. 
Identification, location and language can take various forms depending on the context in which they fit. 
On the Web, the vectors of these three principles that allow anyone to find anything they want online, are the URLs. 

\begin{figure}
\begin{center}
\includegraphics[scale=0.4]{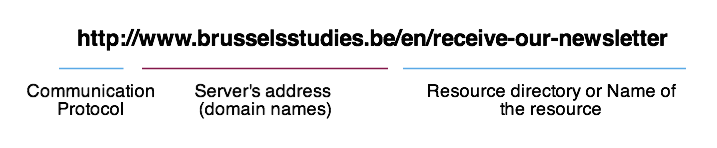}
\caption{\label{url} URL Breakdown}
\end{center}
\end{figure}

Let's take a closer look to the information contained in a URL with the use of an example as showed in Figure \ref{url}.
First, we notice the common communication protocol used on the Web, the \texttt{http} protocol that allows the different components of the networks involved to understand each other by using the same \enquote{language}.
Then, we distinguish the domain name \url{www.brusselsstudies.be} corresponding to the IP address of the wished resource and allowing to locate it on the networks. 
An IP address is a series of numbers\footnote{Or, in the case of the most recent version of the Internet Protocol (IP), IPv6, a series of numbers and letters.} acting as unique identifiers and assigned to each entity connected to the Internet. 
Finally, we notice the resource directory and/or the resource name itself. 

Most of the time we will only retain the part containing the domain names because they are easy to remember and because the web navigators 
have made it possible to access whatever we want by their simple use. Originally, domain names had no intention to be used by end users, 
but their semantic value has made them real standards online when the Web came up and opened the Internet to commerce in the early 1990s. 
Why should we use the Domain Name System? By employing the DNS as top-level directory, URLs exploited the global connectivity offered by that 
system on the Internet \citep[p.107]{mueller_ruling_2004}. Indeed, the DNS was introduced at the beginning of 1980 in order to simplify the 
use and the management of IP adresses. As we have just seen, an IP address is a series of numbers acting as unique identifier on the Internet.

\subsection{From IP addresses to Domain Name System (DNS)}
Like the Internet on the one hand uses the TCP/IP protocols to communicate and has its unique identifiers, namely Internet Protocol addresses (or IP addresses), the Web on the other hand uses the \texttt{http} protocol to interact and URLs to identify and locate documents or other resources online. 
As we know, the Web of Tim Berners-Lee is an application that makes use of the exchange structure that offers the Internet and was built on top of it. 
When the first networks were developed in the 1960s by the Advanced Research Projects Institute (ARPA), the different entities composing these 
networks had to be identified so they could find each other easily in order to communicate. In the beginning there were few hosts on the networks 
which made the identification process and the attribution of IP addresses easy. After a while, it was decided to associate names to these addresses 
composed of numbers for two reasons. First, names serve as mnemonic tools. It is clear that you will rather remember \url{www.google.com} than its corresponding IP address \url{74.125.195.94}. 
Second, the attribution of names makes it easier to modify the addresses. When you change an address you don't need to communicate about it because its corresponding name remains the same \citep[p.195]{weinberg_icann_2000-1}. 

At the time, it was a Network Information Center that used to maintain a text file (e.g. \texttt{host.txt}), composed of translation tables of the IP addresses and their corresponding names.
This translation task was simplified with the emerging of name servers on the different networks. 
In fact, these name servers are in charge to translate the numeric IP addresses into the corresponding human-readable names and vice versa. 
When a host is looking for some information, they will be able to find the resource by simply querying a name server on the network. 
If the name server doesn't have the searched information, it will request another name server and so on until the querying can be answered. 

Over time, the increase in size of the networks and the growing number of hosts engendered three problems regarding organization, 
flexibility and management of the names and their corresponding IP addresses. There was a need to structure the expanding number of data, 
to divide their administration on more servers in order to decrease the weight of downloads and finally to delegate the administration of 
the name servers due to the growing complexity of their management.  

This is how in 1983 scientists of the Information Science Institute of the University of Southern California, including Jon Postel and Paul Mockapetris, 
designed the Domain Name System (DNS). 
The DNS is an optimized implementation of the concept of name server on the Internet, based on a hierarchical structure of different \enquote{zones} corresponding to different portion of the \enquote{namespace} \citep[p.4]{aitchison_pro_2011-2}.

\begin{figure}
\begin{center}
\includegraphics[scale=0.5]{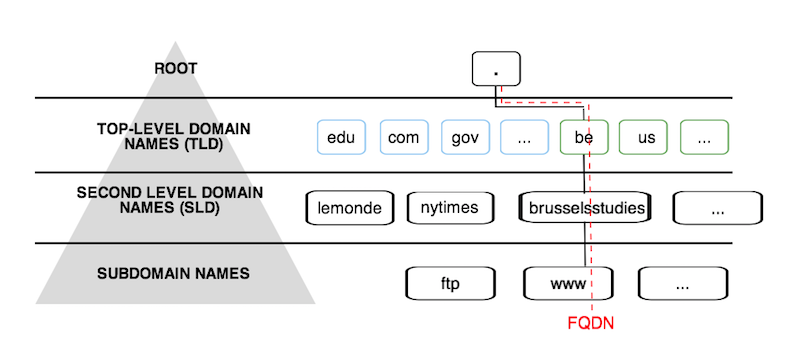}
\caption{\label{dns} DNS Hierarchy}
\end{center}
\end{figure}

The first zone on the top of the hierarchy of the DNS is called the root zone, as showed on top of Figure \ref{dns}.
It is represented by a point, often referred as the \enquote{silent dot} (because we usually don't refer it). This first zone is followed by a second zone which contains the Top-Level Domains (TLD). 
These Top Level-Domain names are divided in two groups: generic Top Level Domains or gTLDs (the first gTLD were .com, .org, .int, .edu, 
.mil, .gov, .net and .arpa) and the country code Top-Level Domains or ccTLDs (like .be, .fr, .us, .jp, etc.). The first group originally designates 
broader categories or groupings like the .edu refers to educative activities and the .com to commercial ones. In the second group, which is based on 
geographical boundaries, we find two-letter codes for every country in the world. 
The second zone is followed by a third one consisting of Second Level Domains (SLD) and so on. If Top-Level Domains are restricted to a certain amount, 
the choice of second level domains is limitless. Each combination though is unique and has an exclusive semantic value. Hence, their craving!

\subsection{Who is in charge?}
The authority in charge of the root zone supervises which Top-Level Domains are visible in the namespace. 
In fact, each \enquote{name} is assigned to an authority which may in turn delegate the authority of lower entities (names) to hers. 
This means that the entity in charge of the root zone controls the whole system. 

In the beginning it was Jon Postel, under contract with the US government, who was responsible of the root. The whole administration of the DNS 
was gathered under the Internet Assigned Names and Number Authority (IANA). The ccTLD where administered to the corresponding countries and the gTLD 
to a few organizations (\enquote{registries}).

Mid 1990, several concerns forced Jon Postel and the US government to rethink the DNS policy.   
First, the authority of the IANA was based on one person. But if something happened to Jon Postel, who would take up the governance of the DNS? 
Second, the emergence of competing entrepreneurs wanting to create alternative namespaces with new Top-Level Domains and independent registries were 
threatening to fragment the whole system \citep[p.201]{klein_icann_2002}. The idea was to reorganize the IANA in a private entity, well-founded, and representative of the 
whole Internet community. 

After a few months of negotiations and adjustments, the American government presented in 1998 the White Paper, birth certificate of the 
Internet Corporation for Assigned Names and Numbers (ICANN). The DNS is controlled by the ICANN ever since.

\subsection{The new generic Top-Level Domains}

There has always been some discussions about how and when new Top Level Domains would be added to the Domain Name System. 
Since the arrival of the Web in the mid 1990s, the DNS has encountered some changes and endowed a kind of economic value \citep[p.105]{mueller_ruling_2004}.
Using them as the top-level directory for the URLs, domain names went from referring network resources to referring content resources. 
They became real benchmarks online. 

Until 2012, the system counted only 22 gTLDs : .com, .arpa, .net, .org, .info, .biz, .mobi, .tel, .name, .asia, .pro, .cat, .jobs, .travel, 
.gov, .coop, .edu, .aero, .museum, .int, .mil and .xxx. 
Since the eight original ones, the ICANN added a few more but remained very strict in the choice of those, although we already can distinguish 
some geographical ones as well as some referring to more specific businesses. 

We had to wait 14 years before ICANN decided to open the system to a massive application period. Through a 185.000\$ application fee everyone 
has the possibility to apply for a Top-Level Domain. The 13th of June 2012, the ICANN announced the entry of 1930 requests. This number was much 
higher then expected. Even if the business was quite discreet on the expected results, they never thought it would encounter such a success.\citep{credou_interview_2016}
On the 1930 applications, more than 751 different candidates were involved. The most popular requests where for the .app, .home, .inc and the .art. 
When a TLDN is coveted by candidates with equal legitimacy two options are offered: or they settle the dispute internal, or an auction is organized and 
the extension will go to the highest bidder \citep[p.39-40]{nazzaro_welcome_2014-2}. Four years later, the applications are still 
processed and new Top Level Domains continue to enter the namespace. The applications came from all over the world: 911 from North America, 
675 from Europe, 303 from Asia-Pacific, 24 from Latin America and the Caribbean and 17 from Africa \citep{icann}.

Belgium also took part of this huge revolution through the addition of three new geographic Top-Level Domains : .brussels, \texttt{.vlaanderen} and \texttt{.gent}. 
The first two TLDs were registered by the registry DNSBelgium (also administering the .be) and the last one was acquired by the registry Combell and the city of Gent. Let us now proceed with a detailed analysis of the uptake of the .brussels TLDN.

\section{Case study: the uptake of the .brussels top-level domain name}
\subsection{Background of the introduction of .brussels}
The story of .brussels started in 2011 when DNS Belgium announced to the three belgian regions (Flemish, Walloons and Brussels Capital) and the german speaking Community, the opportunity to apply for new Top-Level Domains . 

The flemish and the brussels governments went along with the project and made a tender to find a partner who could handle the acquisition process at ICANN and the technical maintenance of the new Top-Level Domains. The non-profit organization, DNS Belgium, got the two of them and entered an application for the .brussels and \texttt{.vlaanderen} in 2012. 

To apply for a new gTLD you had to be an established corporation, organization or institution, pay an evaluation fee of US\$ 185,000 and fulfill the rules of the ICANN's applicant Guidebook. Once the application period was closed, ICANN started the evaluation process. The order of assessment was based on the principle of a lottery. Each candidate had to pick up a passage number. DNS Belgium was quite lucky in its picking and the two extensions were released in June 2014. Where DNS Belgium administer the technical maintenance of the .brussels, it is the city of Brussels who does all the marketing promotion around it. 
The wordplay be.brussels is now fully part of the city's communication and marketing strategy.

\subsection{Harvesting of the research data}
In order to analyse the uptake and usage of the .brussels TLDN, we acquired the full list of all .brussels domains registered until the beginning of March 2016. We requested the data through the Centralized Zone Data Service (CZDS) of ICANN at the end of February. The registry dnsbelgium approved the request on the third of March and released us the entire zone file of its own .brussels extension.
This list, containing 7258 lines, was given to us in an \texttt{*.txt} file.

Some particularities and data quality issues were present in the file. 
After a quick look at the file, we noticed that certain domain names did not make a lot of sense, such as for example 0037thqst4en3so3ahg8c83b6k5bsals.brussels.
These apparent non-sensical URLs are used in the context of the DNSSEC protocol.\footnote{More information can be found on \url{https://ensiwiki.ensimag.fr}.}
We achieved most of the cleaning and data pre-processing using OpenRefine \citep{Packt}\footnote{More information and guidance on \url{http://freeyourmetadata.org}.}, an open-source data transformation tool, as will be described later in this section.

Both a quantitative and a qualitative approach were used to delve into the corpus and analyse its content.
Whilst the quantitative analysis made use of the whole corpus, we resorted to the creation of a representative sample for the manual, qualitative study of the .brussels domain names. 

\subsection{Quantitative approach applied upon the full research corpus}
Using OpenRefine, we have cleaned the corpus and exported a \texttt{*.txt} file containing 5908 domain names, with one domain name per line.

\subsubsection{Percentage of redirections}
If the domain name as such is not being used, it is possible it is \emph{redirected}.
The intention behind verifying redirections is to determine whether .brussels domain names are widely used as main domain names, or if they are mostly used as a \emph{defense mechanism} against domain squatters. 
In order to do just that, we used an existing script\footnote{Available at \href{https://github.com/djenvert/test301}{https://github.com/djenvert/test301}} to test whether an URL is redirected or not. 
Originally, the script takes a \texttt{*.csv} file as input, consisting of two columns : the first one contains the URL that is being tested, and the second one is the URL where the first one should be pointing. 
The script then checks and writes the results in two separate files, one where there is a redirection -- an HTTP 301 status\footnote{https://httpstatuses.com/301} -- and one where there is not.
In our case, since we wanted to verify whether domain names were being redirected or used primarily, the two columns of the input file contained the same content. 
The results of the script is presented in table \ref{redirection}: out of the 5908 domain names tested, 22.7\% were redirected. 
Nonetheless, those results should not be taken at face value: not being redirected does not necessarily mean the domain name is being used, as many of those non-redirects point to nothing (404\footnote{https://httpstatuses.com/404}) or to the confirmation that the domain name has been purchased -- some registrars automatically create a page on newly-bought domain names attesting that the domain name is no longer available. 

\begin{table}[H]
\centering
\begin{tabular}{c|ccc}
        & HTTP 301 & No HTTP 301 & Total \\ \hline
Numbers & 1338     & 4570        & 5908  \\
\%      & 22.7     & 77.3        & 100  
\end{tabular}
\caption{HTTP 301 vs No HTTP 301}
\label{redirection}

\end{table}

\subsubsection{Who owns .brussels domain names?}
Another interesting source of information about domain names is their owners. 
Do we see, as it is the case for real estate, conglomerates speculating on a range of domain names?
In order to determine this, we used \emph{WHOIS}, a protocol -- as defined by Wikipedia -- used to \enquote{[query] databases that store the registered users of assignees of an Internet resource, such as domain name, (...)}.\footnote{\url{https://en.wikipedia.org/wiki/WHOIS}}
For privacy and commercial reasons, most non-commercial tools and websites limit the number of queries a single user can make per day. 
We have found out that the limit for the .brussels TLD, using the latest Debian Jessie \emph{WHOIS} client\footnote{The package is present by default on any Debian installation, and is available at \url{https://packages.debian.org/source/jessie/whois}.}, was of 60 queries per day per IP.
In order to bypass that limitation -- for research, and not commercial purposes -- and correctly query the data for our 5908 domain names, we split our original file into 99 60-domain name files and added the \texttt{whois} command to each line. 
The final files, that were launched from 99 separate IPs, bore the following structure:

\begin{verbatim}
whois -H 0800flowers.brussels
whois -H 10.brussels
whois -H 1000.brussels
whois -H 100masters.brussels
whois -H 100percentbrussel.brussels
\end{verbatim}

The results of the WHOIS querying bore interesting results.
It appears that the most common registrant is the CIRB-CIBG,\footnote{\href{http://cirb.brussels/}{http://cirb.brussels/}} the branch of the Brussels government in charge of introducing new technologies, with a total of 1437 domain names. 
The domain names tend to be \enquote{public service}-oriented, and mostly are names of police zones, postal codes or names of municipalities or names of government branches. 
The second most represented registrant for the .brussels TLD is a private company\footnote{The authors have decided not to disclose the name of the company.} described on its website as \enquote{[being] currently interested in purchasing high quality generic domain names that do not infringe on anyone's intellectual property rights such as grants.com, robot.com, and earth.com}, with 64 unique domain names. 
After those two, the list is composed by Belgian private citizens -- one of them not being from Brussels -- (54, 50 and 44 domain names, respectively), and then by the American multinational company, Apple, with 27 domain names. 
In total, there are 1877 unique registrants. 
If we remove the 1437 domain names of the CIRB-CIBG, it appears that the average number of domain names per registrant is of 1.76. 
This relatively low number seems to indicate that .brussels is not really the victim of large consortium purchasing all domain names.

\subsubsection{Wherefrom are .brussels domain name owners?}
With the personal information available in the \emph{whois} databases, it is also possible to determine the origin of the registrants. 
A breakdown of this data per country of origin is available in table \ref{countries}.

\begin{table}[H]
\centering
\label{countries}
\begin{tabular}{cccc}
\textbf{country of registrant} & \multicolumn{1}{c|}{\textbf{number}} & \multicolumn{1}{l}{\textbf{country of registrant}} & \multicolumn{1}{l}{\textbf{number}} \\ \hline
BE      & \multicolumn{1}{c|}{4200} & \multicolumn{1}{c}{DK}       & \multicolumn{1}{c}{4} \\
US      & \multicolumn{1}{c|}{151}  & \multicolumn{1}{c}{AE}       & \multicolumn{1}{c}{3}   \\
NL      & \multicolumn{1}{c|}{92}   & \multicolumn{1}{c}{IE}       & \multicolumn{1}{c}{3}    \\
DE      & \multicolumn{1}{c|}{84}   & \multicolumn{1}{c}{CY}       & \multicolumn{1}{c}{2}    \\
FR      & \multicolumn{1}{c|}{56}   & \multicolumn{1}{c}{NO}       & \multicolumn{1}{c}{2}    \\
GB      & \multicolumn{1}{c|}{48}   & \multicolumn{1}{c}{VG}       & \multicolumn{1}{c}{2}    \\
CH      & \multicolumn{1}{c|}{16}   & \multicolumn{1}{c}{ZA}       & \multicolumn{1}{c}{1}    \\
LU      & \multicolumn{1}{c|}{14}   & \multicolumn{1}{c}{HU}       & \multicolumn{1}{c}{1}    \\
CA      & \multicolumn{1}{c|}{13}   & \multicolumn{1}{c}{ID}       & \multicolumn{1}{c}{1}    \\
ES      & \multicolumn{1}{c|}{11}   & \multicolumn{1}{c}{IN}       & \multicolumn{1}{c}{1}    \\
AT      & \multicolumn{1}{c|}{10}   & \multicolumn{1}{c}{JP}       & \multicolumn{1}{c}{1}    \\
IL      & \multicolumn{1}{c|}{7}   & \multicolumn{1}{c}{KH}     &  \multicolumn{1}{c}{1}    \\
IT      & \multicolumn{1}{c|}{7}    & \multicolumn{1}{c}{MA}       & \multicolumn{1}{c}{1}    \\
AU      & \multicolumn{1}{c|}{6}    & \multicolumn{1}{c}{PT}       & \multicolumn{1}{c}{1}    \\
CN      & \multicolumn{1}{c|}{6}    & \multicolumn{1}{c}{RO}       & \multicolumn{1}{c}{1}    \\
SE      & \multicolumn{1}{c|}{6}    & \multicolumn{1}{c}{UA}       & \multicolumn{1}{c}{1}    \\
DK      & \multicolumn{1}{c|}{4}    & \multicolumn{1}{c}{(blank)}  & \multicolumn{1}{c}{172} \\
\end{tabular}
\caption{Domain names per country of registrant}
\end{table}

\begin{figure}
\begin{center}
\includegraphics[scale=0.4]{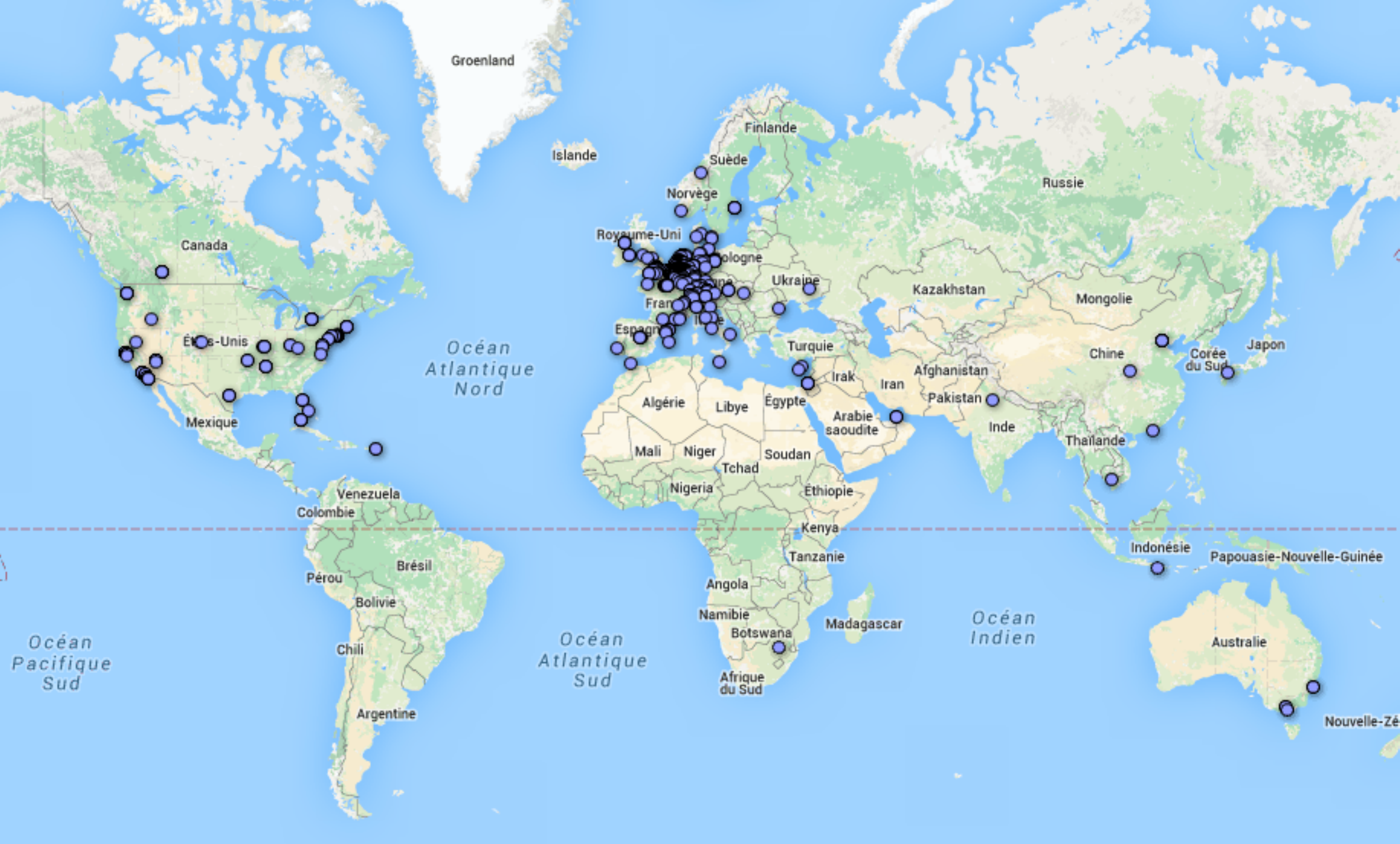}
\caption{\label{map} Visualisation of place names from where .brussels TLDN have been acquired}
\end{center}
\end{figure}

Figure \ref{map} illustrates in a visual manner that the large majority (71,10\%) of the .brussels TLDN have been actually acquired in Belgium, 
even if there are a substantial amount of people elsewhere who also invested in this TLDN.

\subsubsection{How many characters does a .brussels domain name typically contain?}
A last element which can be analysed for the entire corpus is the length repartition of the domain name.
It might seem nonsensical to do so, but a certain number of the new TLDN are often used in order to reduce the length of a domain name.
For example, domain names such as \url{http://del.icio.us/} or
\url{http://flic.kr} integrate the extension as an integral part of the semantics of the URL.
Analysing the average length of the domain names can give an indication of this phenomenon. 

Table \ref{longueur} shows us that more then half of the domain names have a length ranging from ten to five
characters, which can be interpreted as a sign that .brussels is not necessarily purchased to significantly reduce
the length of the domain name.  

\begin{table}[H]
\centering
\label{longueur}
\begin{tabular}{cccc}
length & \multicolumn{1}{c|}{number} & \multicolumn{1}{l}{length} & \multicolumn{1}{l}{number} \\ \hline
8      & \multicolumn{1}{c|}{596}    & 22                         & 31                         \\
7      & \multicolumn{1}{c|}{585}    & 23                         & 25                         \\
6      & \multicolumn{1}{c|}{523}    & 24                         & 20                         \\
9      & \multicolumn{1}{c|}{471}    & 25                         & 16                         \\
10     & \multicolumn{1}{c|}{451}    & 26                         & 9                          \\
5      & \multicolumn{1}{c|}{414}    & 28                         & 9                          \\
4      & \multicolumn{1}{c|}{409}    & 30                         & 9                          \\
11     & \multicolumn{1}{c|}{362}    & 2                          & 8                          \\
3      & \multicolumn{1}{c|}{356}    & 32                         & 7                          \\
12     & \multicolumn{1}{c|}{307}    & 27                         & 6                          \\
13     & \multicolumn{1}{c|}{280}    & 29                         & 5                          \\
14     & \multicolumn{1}{c|}{207}    & 33                         & 3                          \\
15     & \multicolumn{1}{c|}{190}    & 34                         & 2                          \\
17     & \multicolumn{1}{c|}{153}    & 35                         & 2                          \\
16     & \multicolumn{1}{c|}{140}    & 37                         & 2                          \\
18     & \multicolumn{1}{c|}{105}    & 39                         & 2                          \\
19     & \multicolumn{1}{c|}{88}     & 31                         & 1                          \\
20     & \multicolumn{1}{c|}{76}     & 36                         & 1                          \\
21     & \multicolumn{1}{c|}{48}     & 38                         & 1                             
\end{tabular}
\caption{Length repartition of the domain names}
\end{table}

\subsection{Qualitative analysis based on a sample}
In order to gain a more fine-grained understanding of the dataset, we set out to select a sample to be analysed in a
manual way.
Basing our selection on a 99\% confidence score and an interval of 5, a sample of 592 domain names was obtained. 
The selection was carried out randomly, since we feared that variations of the same domain name\footnote{E.g. \url{fullglassrailing.brussels}, \url{full-glassrailing.brussels} and \url{full-glass-railing.brussels}.} would prevent proper representativeness.
To do so, we created a python script\footnote{Available at \url{http://howhotis.brussels}.} which added all domain names to a list, shuffled the list using the \texttt{random} python library and extracted the first 592 domain names of that list. 

Secondly, a set of questions was established in order to guide the analysis. A pragmatic balance had to be found between a sufficient number of questions in order to obtain sufficiently detailed results, but the number of questions also had to be limited in order to keep the manual analysis feasible. A set of eight questions was identified, for which we obtained the following results:
\begin{itemize}
\item \textbf{Parked}: Domain parking \enquote{refers to the registration of an internet domain name without that domain being associated with any services. This may have been done with a view to reserving the domain name for future development, and to protect against the possibility of cybersquatting}.\footnote{\url{https://en.wikipedia.org/wiki/Domain_parking}} Typically, either a message such as \enquote{Under construction} is shown, or advertisements are used to monetise the eyeballs viewing the page.  

For our corpus, \textbf{25,34\%} of the URLs are parked.
\item \textbf{Speculation}: A so-called \emph{domain name aftermarket} has appeared quite quickly after the uptake of the Web in the 1990s. Just like with real-estate, people invest in domain names with the sole purpose of reselling it at a higher price. A typical example here is \url{http://photography.brussels}, which is offered for sale at the price of 9.99\$. We only considered a domain name to fall into this category if the Web page explicitly mentioned the domain name is for sale. Some investors in domain names might decided not to put a specific message online and use other marketing methods to sell the domain name, or just simply wish to wait and see how the market evolves. These approaches are therefore not represented in our results and the figure should therefore be considered as a minimal percentage.   

For our corpus, \textbf{2,87\%} of the domain names are offered in an explicit manner for sale. 

\item \textbf{Language}: Identification of the language used on the Website.
The following languages are represented in our corpus:
\begin{itemize}
\item Multilingual: 159
\item French: 29 
\item English: 23
\item Dutch: 19
\item German: 2
\item Japanese: 1
\item Spanish: 1
\end{itemize}

It comes as no surprise that the majority of websites is offered in more than one language, since the multicultural and -lingual aspect of Brussels.
However, it is to be noted that there are more websites in French than in English or Dutch.  

\item \textbf{Redirection}: This happens when a Webpage has no content of itself, but just forwards you to another URL. This is the case with the example of \url{http://art-nouveau.brussels}, which redirects the user to \url{http://patrimonium.brussels}.
Redirection is mainly used to lower the impact of synonymy and alternate spellings upon users by offering them multiple entry points.

For our corpus, \textbf{27,20\%} of the domain names are redirected towards another URL.

\item \textbf{Sector}: Identification of the sector of activity of the Webpage.
By analysing in an iterative way the content of the websites, a list of the most prominent sectors of activity has been identified. In the case a certain subcategory of an activity is prominent in its own right, a class of its own was created, which happened in the case of culture, tourism, transport (or even politics) which can be considered as a subactivity of services. 
\begin{itemize}
\item \textbf{Services (51,28\%)}: more than half of the websites offer services, including hotel and catering trade, education and medical services for example.
\item \textbf{Commerce (18,38\%)}: almost one fifth offers products for sale.
\item \textbf{Culture (11,54\%)}: more than a tenth of the websites promote cultural activities. 
\item \textbf{Transport (10,68\%)}: this might seem surprising, but it is mainly due to the high number of domain names bought by the STIB.
\item \textbf{Tourism (3,84\%)}: websites promoting tourist events within Brussels.
\item \textbf{Personal websites (2,14\%)}: personal websites of individuals, representing their work or hobby activities.
\item \textbf{Politics (2,14\%)}: websites to promote political parties or individuals.
\end{itemize}
\item \textbf{Public}: \textbf{18,58\%} of the domain names are explicitly associated with a public service.
\item \textbf{Brussels}: \textbf{28,21\%} of the content of the websites is directly or indirectly associated with the city or the region of Brussels.
\end{itemize}

Table \ref{tbl:comp_lang} gives some examples are given of how these questions have been used to analyse the sample corpus.

\begin{table}[H]
\begin{center}
\small
\begin{tabular}{lccccccc}
\hline
\textbf{URL} & \textbf{Parked} & \textbf{Speculation} & \textbf{Language} & \textbf{Redirection} & \textbf{Sector} & \textbf{Public} & \textbf{Bxl}\\
\hline
87seconds & No & No & Multi & Yes & Services & No & Yes \\
photography & No & Yes & EN & No & Services & No & No \\ 
architects & No & No & Multi & No & Services & No & Yes \\
bdebaets & No & No & No info & No & Politics & No & Yes \\
art-nouveau & No & No & Multi & Yes & Culture & Yes & Yes\\
\hline
\end{tabular}
\caption{\label{tbl:comp_lang} Some examples of the domain names from the sample}
\end{center}
\end{table}

\section{Discussion}
\label{ccl}
Let us come back to our initial question: how hot is .brussels in fact? Recent articles in the press have clearly given a negative answer to the question.
The Libre Belgique headed \enquote{Limited success for the .brussels domain name} and the city journal Bruzz \enquote{The .brussels domain name still no success}.\footnote{See \url{http://www.lalibre.be/regions/bruxelles/succes-mitige-pour-le-nom-de-domaine-brussels-57507f5735708ea2d6189236} and \url{http://www.bruzz.be/nl/actua/domeinnaam-brussels-nog-geen-succes}.} DNS Belgium expressed at the launch of .brussels estimates of selling over a period of 5 to 10 years 50.000 domain names. The current uptake (= 4.634 registrations in 2015) remains below those expectations. 

Within this article, we would like to turn the interpretation of the pure figures 180 degrees around and develop a less commercial view on the \emph{hotness} of .brussels.
One of the key outcomes of this article is that the .brussels domain name remains very much a niche player.
However, as citizens of the Internet and lovers of Brussels, is this actually a bad thing?
Within the press, the interpretation of the limited uptake of .brussels is interpreted as a failure, by using the
unique point of view of DNS Belgium, the seller of the new domain names. 
However, we should realise that one can interpret the uptake of .brussels from different angles. 

In order to make a case for our argument, let us refer to the metaphor of the real estate industry.
If there are only few transactions in a certain part of a city, real estate professionals will most certainly deplore the lack of activity and
regret that there's no massive arrival of buyers, who more often than not are short term speculators.
Individual families who buy for example a historic house and put effort into its renovation and contribute in the long term to
the development of rich community and neighbourhood life, don't represent that many transactions and do not necessarily have 
a big impact on the global statistics of the real-estate industry. 
However, from a city planning and long term perspective, these little individual investments can over time develop into small hubs which
contribute significantly to the quality of life in a neighbourhood.    

In this sense, we would like to turn the negative arguments in the press, which only focus on the economical impact of the lagging domain name sells for DNS Belgium, to the value of .brussels as a niche player which has not fallen prey to speculation. 
Even if less .brussels domain names have been acquired than expected, it is interesting to try to discover tendencies in who decided to either put up a new Website or relocate an existing one under the .brussels. The last section gave an overview of the results for a limited set of questions we asked both the entire corpus of registered domain names with the new TLDN .brussels by using computational methods and a sample set, allowing us to delve a bit deeper with manual analyses.

A relatively large portion of the domain names tends to be \enquote{public service}-oriented, and consists mainly of police zones, postal codes or names of municipalities or names of government branches. Specific players active as investors have definitively acquired to some extend .brussels domain names,
but a large portion consiists of Belgian private citizens. If we remove the 1.437 domain names of the CIRB-CIBG, it appears that the average number of domain names per registrant is of 1.76. This relatively low number seems to indicate that .brussels is not really the victim of large consortium purchasing in bulk domain names. This tendency is also reflected in the fact that 71,10 \% of the .brussels domain names have been registered by persons or organisations based in Belgium.

As Thierry Brunfaut, director of the marketing campaign of be.Brussels, mentions on the project website, \enquote{logos, however pretty, are no longer sufficient to build a brand. What makes brands vibrant and strong is their capacity to engage users and build strong connections}\footnote{See the description of the marketing campaign on \url{https://basedesign.com/case-study/be-brussels}.}. This is a strong case of how the .brussels TLDN has been used to complement and boost a traditional visual marketing identity. 
There is an important interaction between the design and marketing of a website, and how .brussels is being operationalised.
For example, \url{http://urbanisme.brussels} is for the moment being redirected towards \url{http://urbanisme.irisnet.be}, but the graphic identity of this website already refers to the .brussels which implies that a future move is in the making.
As the owner of a bed \& breakfast place who has bought about 20 domain names mentioned in an interview: \enquote{.be and .com don't mean anything, whereas the usage of .brussels is immediately tied to the city}. One could consider .brussels as a valuable extra layer of metadata which makes it clear to the end-user that the website is related to the city or region of Brussels. Not only city marketing but also activism immediately linked to the territory of Brussels, such as for example \url{http://mortsdelarue.brussels/} illustrate the power of the .brussels to make the link with the city explicit in the domain name. 

We can therefore end with a positive message: the amount of Starbucks and H\&M shops have remained limited, leaving sufficient room for small and local players with a clear connection to Brussels to acquire a domain name with strong semantics. This is good news both for the responsable of the Website as well as for the end-user. However, this does not mean that the uptake of .brussels should not significantly increase over the years. When interviewing owners of a .brussels, the recurring comment was that end-users are often not aware of the existence of .brussels and even think there is an error in the URL when confronted the first time with an URL ending on .brussels. Promotion and evangelisation should therefore certainly take place over the next years, but we think it is a good thing that the uptake happens in a slow and organic manner instead of being confronted with a \textit{run} on large numbers of domain names, which would just reflect speculative behaviour.  
%\newpage
\small
\bibliographystyle{apalike}
\bibliography{biblio}

\end{document}